\documentclass[12pt]{article}%
\usepackage{amsmath}
\usepackage{amsfonts}
\usepackage{amssymb}
\usepackage{graphicx}%
\providecommand{\U}[1]{\protect\rule{.1in}{.1in}}
\makeatletter
\AtBeginDocument{\@ifpackageloaded{natbib}{\ifNAT@numbers\if@filesw\immediate\write\@auxout{\string\global\string\NAT@numberstrue}\fi\fi}{}}
\makeatother
\begin{document}
\begin{titlepage}
\ \\
\begin{center}
\LARGE
{\bf
The Fall of the Black Hole Firewall:\\
Natural Nonmaximal Entanglement\\
for the Page Curve
}
\end{center}
\ \\
\begin{center}
\large{
Masahiro Hotta
}\\
{\it
Department of Physics, Faculty of Science, Tohoku University,\\
Sendai 980-8578, Japan
}\\
\ \\
\large{Ayumu Sugita}\\
{\it
Department of Applied Physics, Osaka City University,
3-3-138 Sugimoto, Sumiyoshi-ku, Osaka 558-8585, Japan }
\end{center}
\begin{abstract}
The black hole firewall conjecture is based on the Page curve hypothesis, which claims that entanglement between a black hole and its Hawking radiation is almost maximum.
Adopting canonical typicality for nondegenerate systems with nonvanishing Hamiltonians, we show the entanglement becomes nonmaximal, and energetic singularities (firewalls) do not emerge for general systems.
An evapolating old black hole must evolve in Gibbs states with exponentially small error probability after the Page time as long as the states are typical. This means that ordinary used microcanonical states are far from typical.
The heat capacity  computed from the Gibbs states should be nonnegative in general.
However the black hole heat capacity is acutally negative due to the gravitational instability.
Consequently the states are not typical until the last burst.
This requires inevitable modification of the Page curve, which is based on the typicality argument.
For static thermal pure states of a large AdS black hole and its Hawking radiation,  the entanglement entropy equals the thermal entropy of the smaller system.
\end{abstract}
\end{titlepage}

\section{Introduction}

\ \ 

\ The interesting possibility of black hole firewalls was proposed from a
viewpoint of quantum information \cite{firewall} and has attracted much
attention. In the firewall conjecture, black hole horizons are not a smooth
region even for free fall observers who attempt to pass through it. On the
horizon the observers see highly energetic quantum walls (firewalls) before
they collide against it and burn up.

Essentially the firewall conjecture is based on the Page curve hypothesis of
black hole evaporation \cite{pagetime}, and the hypothesis comes from the
Lubkin-Lloyd-Pagels-Page theorem (LLPP theorem) \cite{l}\cite{LP}\cite{page}.
According to the LLPP theorem, quantum entanglement between two macroscopic
systems $S_{I}$ and $S_{II}$ is almost maximum in a typical pure state
$|\Psi\rangle_{I,II}$ of the composite Hilbert space $\mathcal{H}_{I}%
\otimes\mathcal{H}_{II}$, assuming that dimension $N_{II}$ of $\mathcal{H}%
_{II}$ is much larger than dimension $N_{I}$ of $\mathcal{H}_{I}$. The reduced
density operator (quantum state) $\hat{\rho}_{I}=\operatorname*{Tr}%
_{II}\left[  |\Psi\rangle_{I,II}\langle\Psi|_{I,II}\right]  $ of $S_{I}$
almost equals $\hat{I}/N_{I}$, where $\hat{I}$ is unit matrix acting on
$\mathcal{H}_{I}$. Inspired by this theorem, Page came up with a fascinating
scenario for information leakage of evaporating black holes. He thinks that
evaporation of a macroscopic black hole in an initial pure state is modeled by
two quantum systems $B$ and $R$ with finite, but time dependent dimensions
$N_{B}$ and $N_{R}$. $B$ represents internal degrees of freedom of the black
hole, and $R$ represents the Hawking radiation out of the black hole. It may
be possible that the finiteness of $N_{B}$ and $N_{R}$ is justified if quantum
gravity is taken account of. Such a quantum effect may truncate the degrees of
freedom in a higher energy scale than Planck energy, like string theory. In
condensed matter physics, total energy $E$ is quite low. Thus high-energy
density of states around the cutoff scale of the system becomes irrelevant. So
it is enough to treat a finite dimensional Hilbert space to describe the physics.

The essence of Page's hypothesis is summarized into the following propositions
for entanglement between $B$ and $R$:

\bigskip

(I) When $N_{R}\gg N_{B}$ (or $N_{R}\ll N_{B}$), $B$ and $R$ in a typical pure
state of quantum gravity share almost maximal entanglement. In other words, a
typical quantum state of the smaller system among $B$ and $R$ is almost
proportional to unit matrix $\hat{I}$.

\bigskip

(II) Entanglement entropy $S_{EE}$ of the smaller system among $B$ and $R$ is
equal to its thermal coarse-grained entropy.

\bigskip

\bigskip\begin{figure}[ptb]
\begin{center}
\includegraphics[width=12cm]{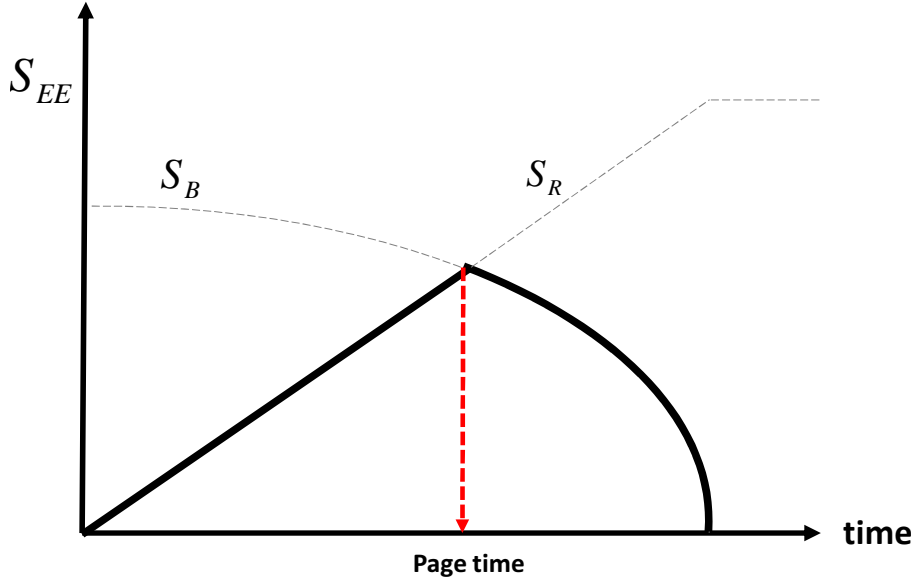}
\end{center}
\caption{schematic figure of Page curve. In the conjecture, entanglement
entropy between black hole and Hawking radiation equals thermal entropy of the
smaller system and attains almost maximum value at each time.}%
\end{figure}Proposition (I) is clearly motivated by the LLPP theorem.
Combining (I) and (II), it is deduced that $S_{EE}$ between $B$ and $R$ takes
almost the maximum value and equals the Bekenstein-Hawking entropy of black
holes $S_{B}=\mathcal{A}/(4G)$ after the Page time, at which decreasing
$S_{B}$ equals increasing thermal entropy $S_{R}$ of the Hawking radiation.
The Page time is estimated as about 53\% of the lifetime of evaporating black
holes, and the mass at the Page time is about 77\% of initial mass
\cite{pagetime}. Thus the black hole remains macroscopic at the Page time, and
its semi-classical picture is valid. Black holes after the Page time are
referred to as \textquotedblleft old\textquotedblright. Before the Page time,
$S_{EE}$ is equal to $S_{R}$, and the black holes are referred to as
\textquotedblleft young\textquotedblright. \ Since time evolution of $S_{B}$
and $S_{R}$ is computed in an established semi-classical way, this argument
provides a prediction for the time curve of $S_{EE}$ during the evaporation.
This is the Page curve. Its schematic figure is given in figure 1.

The firewall conjecture arises basically from (I). For an old black hole,
Hawking radiation $R$ is decomposed into $A$, which is emitted after the Page
time, and $C$, which is emitted before the Page time. This is depicted in
figure 2 for gravitational collapse of a massless shell. The dimensions of
sub-Hilbert spaces for $A$ and $C$ are denoted by $N_{A}$ and $N_{C}$. Due to
the old age of the black hole, $N_{C}\gg N_{A}N_{B}$ is satisfied. From (I),
$AB$ system is almost maximally entangled with $C$. Thus a typical quantum
state of $AB$ can be approximated as $\hat{\rho}_{AB}\approx\hat{I}%
_{AB}/(N_{A}N_{B})$. Since the unit matrix $\hat{I}_{AB}$ is written
as\ $\hat{I}_{A}\otimes\hat{I}_{B}$, no correlation exists between $A$ and $B$
\cite{HH}\cite{harlow}. Consequently, for example, kinetic energy terms of
quantum fields for the Hawking radiation diverge on the horizon. Let us denote
an inside point $x_{B}$ near the horizon, and an outside point $x_{A}$. Then a
kinetic term $\left(  \partial_{x}\hat{\varphi}\right)  ^{2}~$of a scalar
field $\hat{\varphi}(x)~$is given by $\left(  \hat{\varphi}(x_{A}%
)-\hat{\varphi}(x_{B})\right)  ^{2}/\epsilon^{2}$, where $\epsilon$ is
ultraviolet cutoff (lattice spacing). Apparently, when $\epsilon\rightarrow0$,
this diverges like $1/\epsilon^{2}$ on the horizon for the typical state
$\hat{\rho}_{AB}\varpropto\hat{I}_{A}\otimes\hat{I}_{B}$ because of the
correlation loss, and a firewall emerges. \begin{figure}[ptb]
\begin{center}
\includegraphics[width=6cm]{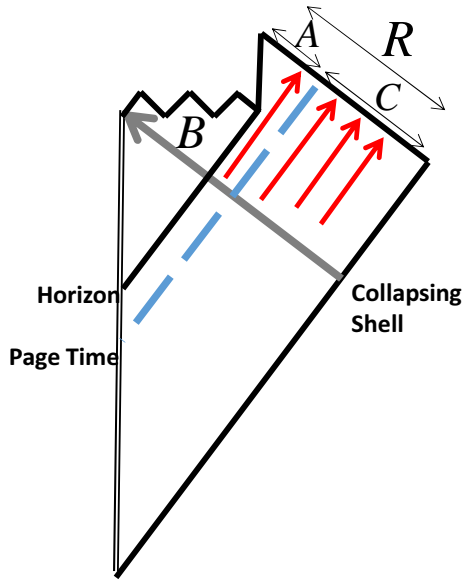}
\end{center}
\caption{early Hawking radiation emitted before Page time and late radiation
emitted out of an old black hole after Page time.}%
\end{figure}Besides the strong subadditivity paradox \cite{M}\cite{firewall}
is often worried about in the context of the firewall paradox. Let us suppose
a strong subadditivity inequality of the von Neumann entropy for an old black
hole $A$, late radation $B$ and early radiation $C$:%
\begin{equation}
S_{AB}+S_{AC}\geq S_{A}+S_{ABC}. \label{0}%
\end{equation}
Assuming the no drama conditions $S_{AB}=0$ and $S_{ABC}=S_{C}$ in the
references \cite{M} and \cite{firewall} yields%
\[
S_{AC}\geq S_{A}+S_{C}.
\]
As long as the old black hole continuously emits the stored information after
the Page time, the purity of $AC$ system increases, and $S_{AC}$ decreases in
time. Thus $S_{A}>S_{AC}$ holds. This leads to an apparent contradiction that
$S_{C}\leq0$. Actually the early radiation has positive thermal entropy
$S_{C}>0$. This seems to mean a breakdown of the no drama condition and
suggests the existence of firewalls. However it is already known that the
paradox is clearly avoided in moving mirror models \cite{HMF} and long-lived
remnant models \cite{AS}. In \cite{HMF}, it is pointed out that spacial
locality among the subsystems $A,B$ and $C$ is ill-defined. Consequently the
$AB$ system inevitably has nonvanishing entanglement with zero-point
fluctuation of the radiation field, and $S_{AB}=0$ does not hold even if we
postulate the no drama condition for the horizon in a physical sense. This
remains true in black hole evaporation. For the long-lived remnant models,
$S_{AC}$ does not decrease even after the Page time, but rather increases
until the last burst of the evaporating black hole. Thus $S_{A}>S_{AC}$ is not
satisfied and the paradox is evaded. Similarly, any evaporation scenario, in
which no information is emitted out of the black hole until the last burst, is
free of the strong subadditibity paradox at least. Though the strong
subadditivity paradox can be avoided, the long-lived remnant models
\cite{remnant1}\cite{remnant2}\cite{remnant3} are supposed to have other flaws
\cite{preskill}. The energy of the remnant is of order of the planck mass, but
in order to store the huge amount of information, the remnants seem to possess
almost infinite degenaracy. The tremendous degeneracy may break the past great
success of many experiments and observations via loop effects in particle
scattering processes and species summation in partition functions for thermal
equilibrium in early universe \cite{remnant3}\cite{preskill}. \ In this paper,
in order to avoid those flaws and firewalls simultaneously, we consider an
alternative scenario. In the scenario, all the information come out at the
last burst. Of course, the total energy of the last ray out of the black hole
is merely of order of Planck mass and very tiny. However, as stressed first by
Wilczek \cite{w}, an outgoing zero-point fluctuation flow of quantum fields,
which is adjacent to the last ray, can share the huge amount of entanglement
between the Hawking radiation emitted before. This fluctuation flow in a local
vacuum region has zero energy, but transports the information to the future
null infinity without any contradiction. At the last burst, quantum gravity
critically affects the horizon. Hence the no drama condition is no longer
required. Thus we do not need to care the strong subadditiviy paradox even if
the entropy $S_{AC}$ suddenly decreases at the last burst.

In this paper, first of all, it is pointed out that proposition (I) does not
hold if nondegeneracy of energy eigenstates of the total system is taken
account of. The typical states have to be exponentially close to Gibbs states
with finite temperatures. The entanglement between $AB$ and $C$ becomes
nonmaximal. Therefore, without breaking monogamy of entanglement, $A$ is able
to share entanglement with $B$, and simultaneously with $C$. The entanglement
between $A$ and $B$ yields a correlation that makes the horizon smooth, and no
firewall appears. Though it has been proven that such nonmaximal entanglement
prevents the emergence of firewalls in moving mirror models \cite{HMF}, more
stringent arguments are provided for general systems in this paper. Our result
means that the ordinary used microcanonical states in the arguments of
\cite{firewall} and  \cite{pagetime} are far from typical for quantum
entanglement between a black hole and its Hawking radiation. In section 2, we
briefly review a general formulation of canonical typicality with nonzero
Hamiltonians \cite{neumann, s, goldstein, popescu, reimann}. Our discussion is
based on \cite{s}. From the rigorous results, it turns out that proposition
(I) does not hold for general systems which satisfy natural conditions.
Therefore it turns out that, after the Page time, the evapolating old black
hole must evolve in Gibbs states with high precision as long as the pure state
of the black hole and the Hawking radiation is typical. In general, heat
capacity computed from a partition function $Z(1/T)$ of a Gibbs state must be
nonnegative as%
\[
\frac{d\left\langle \hat{H}\right\rangle }{dT}=\frac{1}{T^{2}}\left\langle
\left(  \hat{H}-\left\langle \hat{H}\right\rangle \right)  ^{2}\right\rangle
\geq0,
\]
where $\hat{H}$ is the Hamiltonian of the system, $T$ is temperature and
$\left\langle \cdot\right\rangle =\operatorname*{Tr}\left[  \cdot\exp\left(
-\hat{H}/T\right)  \right]  /Z(1/T)$. However the black hole heat capacity is
acutally negative due to the gravitational instability. For instance,
Schwarzschild blacl holes, the energy $E$ is its mass $M~$and equals $\left(
8\pi GT\right)  ^{-1}$. The heat capacity is computed as negative: $\frac
{dE}{dT}=-\left(  8\pi GT^{2}\right)  ^{-1}<0$. Thus the pure state of the
system is never typical until the last burst. This leads to inevitable
modification of the Page curve. In section 3, we discuss proposition (II). In
black hole evaporation, the proposition is implausible. From a viewpoint of
semi-classical general relativity, it looks more fascinating to take an
alternative for the Page curve. The entanglment entropy continues to increase
even after the Page time. At the last burst, it suddenly goes to zero and all
the information is retrieved. Finally, it is commented that typical
entanglement entropy of a large AdS black hole and its Hawking radiation
equals thermal entropy of the smaller system. We adopt natural unit,
$c=\hbar=k_{B}=1$.

\bigskip

\section{Nonmaximality of Entanglement in Canonical Typicality}

\bigskip

In this section, we claim that proposition (I) is not satisfied for general
systems with nondegenerate Hamiltonian. When pure states are randomly sampled
in a sub-Hilbert space with fixed total energy, a typical state is not
maximally entangled. Then the corresponding state of the smaller subsystem is
not the completely mixed state which is proportional to the unit matrix, but a
Gibbs state. Let us think two finite quantum systems $S_{1}$ and $S_{2}$,
whose dimensions of Hilbert spaces $\mathcal{H}_{1}$ and $\mathcal{H}_{2}$ are
denoted by $D_{1}$ and $D_{2}$, respectively. Let us think a pure state
$|\Psi\rangle_{12}$ in $\mathcal{H}_{1}\otimes\mathcal{H}_{2}$. Its density
operator $\hat{\rho}_{12}=|\Psi\rangle_{12}\langle\Psi|_{12}$ is a $D_{1}%
D_{2}\times D_{1}D_{2}$ Hermitian matrix. Thus it can be expanded uniquely in
terms of a basis of $U(D_{1}D_{2})$ Hermitian generators $\left\{  \hat
{I}\otimes\hat{I},\hat{G}_{n\mu}\right\}  $:%
\[
\hat{\rho}_{12}=\frac{1}{D_{1}D_{2}}\left(  \hat{I}\otimes\hat{I}+\sum_{n\mu
}\left\langle \hat{G}_{n\mu}\right\rangle \hat{G}_{n\mu}\right)  ,
\]
where $\hat{G}_{n\mu}$ are traceless and satisfy $\operatorname*{Tr}\left[
\hat{G}_{n\mu}\hat{G}_{n^{\prime}\mu^{\prime}}\right]  =D_{1}D_{2}%
\delta_{nn^{\prime}}$ and
\[
\left\langle \hat{G}_{n\mu}\right\rangle =\operatorname*{Tr}\left[  \hat{\rho
}_{12}\hat{G}_{n\mu}\right]  =\langle\Psi|_{12}\hat{G}_{n\mu}|\Psi\rangle
_{12}.
\]
The set of $\hat{G}_{n\mu}$ consist of basis generators $\hat{T}_{n}$ and
$\hat{R}_{\mu}$ for each sub-Hilbert space. $\hat{T}_{n}$ and $\hat{R}_{\mu}$
are traceless and Hermitian, and obey the following normalization:
\begin{align*}
\operatorname*{Tr}_{1}\left[  \hat{T}_{n}\hat{T}_{n^{\prime}}\right]   &
=D_{1}\delta_{nn^{\prime}},\\
\operatorname*{Tr}_{2}\left[  \hat{R}_{\mu}\hat{R}_{\mu^{\prime}}\right]   &
=D_{2}\delta_{\mu\mu^{\prime}}.
\end{align*}
For $n=1\sim D_{1}^{2}-1$, $\hat{G}_{n0}$ is defined as
\[
\hat{G}_{n0}=\hat{T}_{n}\otimes\hat{I}.
\]
Similarly, for $\mu=1\sim D_{2}^{2}-1$, $\hat{G}_{0\mu}$ is defined as
\[
\hat{G}_{0\mu}=\hat{I}\otimes\hat{R}_{\mu}.
\]
The remaining generators are given by%
\[
\hat{G}_{n\mu}=\hat{G}_{n0}\hat{G}_{0\mu}=\hat{T}_{n}\otimes\hat{R}_{\mu}%
\]
The reduced quantum state $\hat{\rho}_{1}$ for $S_{1}$ is computed as%

\[
\hat{\rho}_{1}=\operatorname*{Tr}_{2}\left[  |\Psi\rangle_{12}\langle
\Psi|_{12}\right]  .
\]
It is worth noting that $\hat{\rho}_{1}$ is uniquely fixed by measuring only
$D_{1}^{2}-1$ expectation values of $\hat{G}_{n0}$ with respect to
$|\Psi\rangle_{12}$. This is because
\[
\left\langle \hat{G}_{n0}\right\rangle =\langle\Psi|_{12}\hat{G}_{n0}%
|\Psi\rangle_{12}=\langle\Psi|_{12}\left(  \hat{T}_{n}\otimes\hat{I}\right)
|\Psi\rangle_{12}=\operatorname*{Tr}_{1}\left[  \hat{\rho}_{1}\hat{T}%
_{n}\right]  =\left\langle \hat{T}_{n}\right\rangle
\]
holds. In fact, $\hat{\rho}_{1}$ is written as
\[
\hat{\rho}_{1}=\frac{1}{D_{1}}\left(  \hat{I}+\sum_{n}\left\langle \hat{T}%
_{n}\right\rangle \hat{T}_{n}\right)  =\frac{1}{D_{1}}\left(  \hat{I}+\sum
_{n}\left\langle \hat{G}_{n0}\right\rangle \hat{T}_{n}\right)  .
\]
In order to analyze the smaller system $S_{1}$, expectation values of other
components of $\hat{G}_{n\mu}$ are not required.

\bigskip\begin{figure}[ptb]
\begin{center}
\includegraphics[width=12cm]{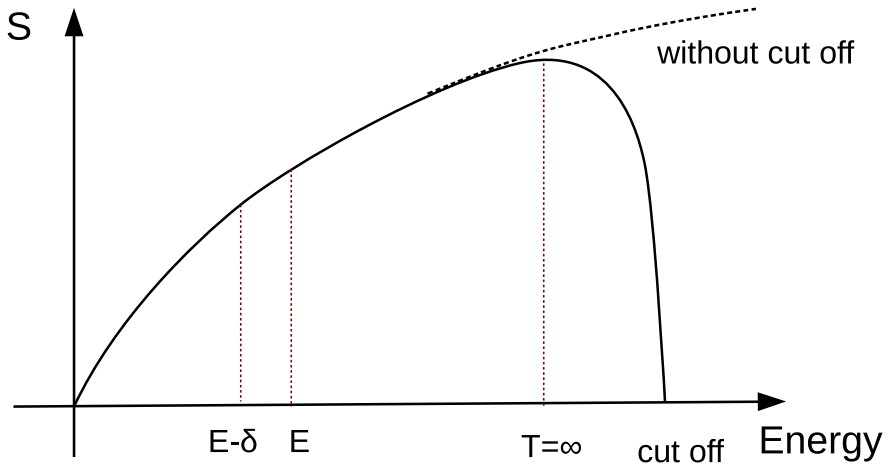}
\end{center}
\caption{behavior of entropy in ordinary systems. When we focus on an energy
shell $\left[  E-\delta,E\right]  $, the cutoff dependence becomes irrelevant
for the canonical typicality argument.}%
\end{figure}Next let us think a nondegenerate Hamiltonian $\hat{H}$ for the
composite system. $\hat{H}$ takes its general form of%

\[
\hat{H}=\hat{H}_{1}\otimes\hat{I}+\hat{I}\otimes\hat{H}_{2}+\hat{V}_{12},
\]
where $\hat{H}_{1}$ and $\hat{H}_{2}$ are free Hamiltonians, and $\hat{V}%
_{12}$ is an interaction term between the subsystems. For simplicity, we
ignore $\otimes$ and $\hat{I}$ in later equations such that $\hat{H}%
_{1}\otimes\hat{I}$ is abbreviated as $\hat{H}_{1}$. The normalized
eigenstates of $\hat{H}$ with eigenvalue $E_{j}$ are denoted by $|E_{j}%
\rangle$:
\[
\hat{H}|E_{j}\rangle=E_{j}|E_{j}\rangle.
\]
We define a set $\Delta(E)$ of energy indices for a macroscopically large
total energy $E$ and a positive number $\delta$ as
\[
\Delta(E)=\left\{  j|E_{j}\in\left[  E-\delta,E\right]  \right\}  .
\]
Let us introduce a sub-Hilbert space $\mathcal{H}_{\Delta(E)}$, which is
spanned by $\left\{  |E_{j}\rangle|j\in\Delta(E)\right\}  $ and its dimension
is denoted by $D$. Microcanonical energy shell is defined as the set of pure
states in $\mathcal{H}_{\Delta(E)}$. It should be stressed that $\mathcal{H}%
_{\Delta(E)}$ is not a tensor product $\mathcal{H}_{B}\otimes\mathcal{H}_{R}$
of any sub-Hilbert spaces $\mathcal{H}_{B}$ of $\mathcal{H}_{1}$ and
$\mathcal{H}_{R}$ of $\mathcal{H}_{2}$. In order to understand this, let us
suppose a case in which $\hat{V}_{12}$ is negiligibly small. Then
$\mathcal{H}_{\Delta(E)}$ is spanned by $\left\{  |E_{1}\rangle_{1}%
|E-E_{1}\rangle_{2}\right\}  $, where $|E_{1}\rangle_{1}$ is eigenstate with
eigenvalue $E_{1}$ of $\hat{H}_{1}$, $|E-E_{1}\rangle_{2}$ is eigenstate with
eigenvalue $E-E_{1}$ of $\hat{H}_{2}$. However, $|E_{1}\rangle_{1}%
|E-E_{1}^{\prime}\rangle_{2}$ with $E_{1}\neq E_{1}^{\prime}$ is not included
by $\mathcal{H}_{\Delta(E)}$. This clearly implies $\mathcal{H}_{\Delta
(E)}\neq\mathcal{H}_{B}\otimes\mathcal{H}_{R}$. Hence, the tensor product
structure assumption of the Page curve hypothesis is not appropriate for
descriptions of black hole evaporation.

For ordinary physical systems with large volume $V$, $D$ becomes exponentially
large like $\exp\left(  \gamma V\right)  $ for with a positive constant
$\gamma$. Taking a small value of $\delta$ gives us a naive picture of for the
energy shell, which often appears in standard textbooks of statistical
mechanics. Note that $\delta$-dependence for final results of statistical
mechanics is irrelevant in general. In fact, $\delta$ does not necessarily
have to be small in the later discussion, since the density of states
$e^{S(E)}$, where $S$ is the entropy, is a very rapidly increasing function
and the eigenstates close to the upper bound $E$ give dominant contribution,
as depicted in figure 3. Therefore, for simplicity, it is also possible to
take an energy shell, say, $\left[  0,E\right]  $ for $\Delta(E)$ instead of
$\left[  E-\delta,E\right]  $.

Any pure state in $\mathcal{H}_{\Delta(E)}$ is written as%

\begin{equation}
|\Psi\rangle_{12}=\sum_{j\in\Delta(E)}c_{j}|E_{j}\rangle, \label{1}%
\end{equation}
where $c_{j}$ satisfy the normalization condition, $\sum_{j\in\Delta
(E)}\left\vert c_{j}\right\vert ^{2}=1$. In order to analyze canonical
typicality for $\mathcal{H}_{\Delta(E)}$, let us introduce a uniform
probability distribution for $c_{j}$ as
\[
p\left(  c_{1},\cdots,c_{D}\right)  =\frac{\Gamma\left(D\right)
}{\pi^{D}}\delta\left(  \sum_{j\in\Delta(E)}\left\vert c_{j}\right\vert
^{2}-1\right)  ,
\]
such that $\int p\left(  c_{1},\cdots,c_{D}\right)  d^{D}c=1$. The ensemble
average value of a function $f$ of $c_{j}$ with respect to a unit sphere of
$\mathcal{H}_{\Delta(E)}~$(microcanonical energy shell) is computed as
\[
\overline{f}=\int f\left(  c_{1},\cdots,c_{D}\right)  p\left(  c_{1}%
,\cdots,c_{D}\right)  d^{D}c.
\]
The ensemble average of a quantum expectation value $\left\langle \hat
{O}\right\rangle $ of an observable $\hat{O}$ in $|\Psi\rangle_{12}$ in eq.
(\ref{1}) is denoted by $\overline{\left\langle \hat{O}\right\rangle }$. The
statistical deviation from $\overline{\left\langle \hat{O}\right\rangle }$ is
given by%

\[
\delta\left\langle \hat{O}\right\rangle =\left\langle \hat{O}\right\rangle
-\overline{\left\langle \hat{O}\right\rangle }.
\]
As proven in Appendix, the ensemble mean square error of $\left\langle \hat
{O}\right\rangle $ is upper-bounded as
\begin{equation}
\overline{\left(  \delta\left\langle \hat{O}\right\rangle \right)  ^{2}%
}=\overline{\left\langle \hat{O}\right\rangle ^{2}}-\overline{\left\langle
\hat{O}\right\rangle }^{2}\leq\frac{\left\Vert \hat{O}^{2}\right\Vert }{D+1},
\label{8}%
\end{equation}
where the operator norm $\left\Vert \hat{O}^{2}\right\Vert $ represents the
maximum absolute value of the eigenvalues of $\hat{O}^{2}$. \ By taking
$\hat{O}=\hat{G}_{n0}$, we have \cite{s}%

\begin{equation}
\overline{\left(  \delta\hat{G}_{n0}\right)  ^{2}}\leq\frac{\left\Vert \hat
{G}_{n0}^{2}\right\Vert }{D+1}.\label{2}%
\end{equation}
It should be stressed that $\left\Vert \hat{G}_{n0}^{2}\right\Vert \left(
=\left\Vert \hat{T}_{n}^{2}\right\Vert \right)  $ is independent of $D_{2}$,
though $D~$grows exponentially as $\exp\left(  \gamma V_{2}(D_{2})\right)  $
with respect to volume $V_{2}(D_{2})$ of $S_{2}$. Hence the right hand side in
eq. (\ref{2}) becomes negligibly small as $\exp\left(  -\gamma V_{2}%
(D_{2})\right)  $ for large $D_{2}$ with $D_{1}$ fixed. Because the
statistical fluctuation is so small, typical values of $\left\langle \hat
{G}_{n0}\right\rangle $ are very close to the central value $\overline
{\left\langle \hat{G}_{n0}\right\rangle }$. This implies that $\hat{\rho}_{1}$
for typical $|\Psi\rangle_{12}$ in $\mathcal{H}_{\Delta(E)}$ coincides with
its ensemble average state $\overline{\hat{\rho}_{1}}$ with almost certainty.
In fact, we are able prove that the ensemble deviation of $\hat{\rho}_{1}$ is
estimated \cite{s} as
\begin{equation}
\overline{\operatorname*{Tr}_{1}\left[  \left(  \hat{\rho}_{1}-\overline
{\hat{\rho}_{1}}\right)  ^{2}\right]  }\leq\frac{1}{D_{1}\left(  D+1\right)
}\sum_{n}\left\Vert \hat{G}_{n0}^{2}\right\Vert ,\label{7}%
\end{equation}
the right hand side of which decays rapidly due to $D$ divergence as $D_{2}$
becomes large. Note that $\overline{\hat{\rho}_{1}}$ is given by
\[
\overline{\hat{\rho}_{1}}=\frac{1}{D_{1}}\left(  \hat{I}+\sum_{n}%
\overline{\left\langle \hat{G}_{n0}\right\rangle }\hat{T}_{n}\right)  .
\]
Due to $\hat{V}_{12}$, energy is exchanged between $S_{1}$ and $S_{2}$ as
shown in figure 4. If the contribution of $\hat{V}_{12}$ is negligibly small
compared to $\hat{H}_{1}$ and $\hat{H}_{2}$, The sum, $\hat{H}_{1}$ $+$
$\hat{H}_{2}$, is approximately conserved. Then, as proven in many textbooks
of statistical mechanics, $\overline{\hat{\rho}_{1}}$ becomes a Gibbs state
with a fixed temperature with high precision for $D_{2}\gg D_{1}$:%
\begin{equation}
\overline{\hat{\rho}_{1}}\approx\frac{1}{Z(\beta)}\exp\left(  -\beta\hat
{H}_{1}\right)  ,\label{3}%
\end{equation}
for ordinary physical systems. The difference between the typical state and
the Gibbs state must be exponentially small: $\overline{\operatorname*{Tr}%
_{1}\left[  \left(  \hat{\rho}_{1}-\overline{\hat{\rho}_{1}}\right)
^{2}\right]  }\leq C\exp\left(  -\gamma V\right)  $. The inverse temperature
$\beta$ is determined by the total energy $E$, which is much less than the
cutoff energy scale for black hole evaporation. Contrary to proposition (I),
eq. (\ref{3}) shows that $\overline{\hat{\rho}_{1}}$ is not proportional to
$\hat{I}_{1}$ if the temperature is finite, unless $\hat{H}_{1}=0$. Note that
a microcanonical state $\hat{\rho}_{m}$, which is proportional to the
projection operator $\hat{I}_{E}$ onto the microcanonical energy shell, is far
from typical, even though its von Neumann entropy $I_{m}$ is $O(V)$ and the
difference from the Gibbs state entropy $I_{c}$ 
is merely $O(\ln V)$. For the von
Neumann entropy $I_{t}$ of a typical state, $\left\vert I_{t}-I_{c}\right\vert
$ is not $O(\ln V)$, but exponentially small as $O(\exp\left(  -\gamma
V\right)  )$. 
\footnote{Note that $\overline{\hat{\rho}_1}$ 
is not exactly the same as the Gibbs state $e^{-\beta \hat{H}_1}/Z$, 
because of a small correction due to the interaction term $\hat{V}_{12}$.
Therefore, precisely speaking, $I_c$ should be regarded as the
von Neumann entropy of $\overline{\hat{\rho}_1}$.}
The entropy difference $\left\vert I_{m}-I_{c}\right\vert =O(\ln V)$
is too large to regard the microcanonical state as a typical state.
Nondegeneracy of $\hat{H}$ provides nonmaximal entanglement between $S_{1}$
and $S_{2}$ to elude the strong subadditivity paradox. It has been pointed out
\cite{HMF} that such a nonmaximal entanglement appears in moving mirror models
and avoids firewalls. The above argument extends the moving mirror result to
general ones. This completely removes the reason for black hole firewall
emergence in reference \cite{firewall}.

For evaporating old black holes, eq. (\ref{3}) implies that state
superposition of different energy black holes emerges in the total pure state
of $B$ and $R$, and generates much more entanglement, compared to a single
black hole contribution with a fixed energy in the Page curve hypothesis. It
should be noted that when the Gibbs state in eq. (\ref{3}) is
thermodynamically unstable due to the emergence of negative heat capacity,
just like for asymptotically flat black hole spacetimes \cite{HP}, typicality
argument itself is unable to be applied to black hole evaporation, and never
provides any correct insight. Quantum states in the evaporation are nontypical
all the time, and proposition (I) loses its reasoning. \begin{figure}[ptb]
\begin{center}
\includegraphics[width=12cm]{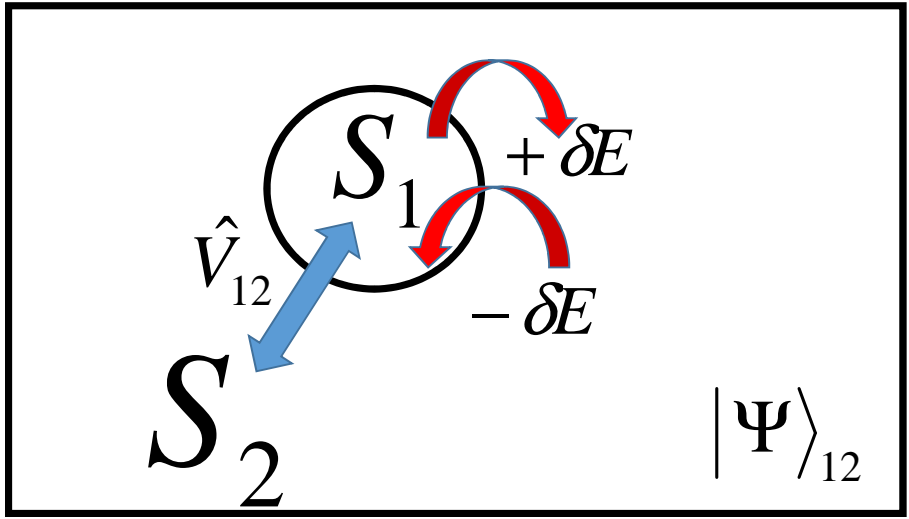}
\end{center}
\caption{a canonical typicality setup for general two systems which preserves
total energy and allows energy transportation between subsystems
bi-directionally.}%
\end{figure}

The result of canonical typicality with eq. (\ref{3}) has been already
commented by Harlow for weak interaction limit \cite{harlow}. However it
should be emphasized that Harlow does not give any proof of the typicality.
Harlow also pointed out \cite{harlow} that there remains a subtlety for the
firewall removal even if we accept the canonical typicality. Let us recall the
setup of gravitational collapse in figure 2. Naively, it may be expected that
the quantum state $\overline{\hat{\rho}_{AB}}$ of late radiation $A$ and black
hole $B$ for a typical state is approximated by%

\begin{equation}
\frac{1}{Z(\beta)}\exp\left(  -\beta\left(  \hat{H}_{A}+\hat{H}_{B}\right)
\right)  =\frac{1}{Z(\beta)}\exp\left(  -\beta\hat{H}_{A}\right)  \otimes
\exp\left(  -\beta\hat{H}_{B}\right)  \label{4}%
\end{equation}
The above tensor product structure of the state means no correlation between
$A$ and $B$ even after taking account of $\hat{H}$ nondegeneracy. Therefore,
just as the original story of firewalls, the expectation value of the kinetic
energy density of radiation fields might be divergent. However, we claim that
the worry is useless. In order to see no divergence, we go back to the above
general formulation. The system Hamiltonian is given by
\[
\hat{H}=\hat{H}_{A}+\hat{H}_{B}+\hat{H}_{C}+\hat{V}_{AB}+\hat{V}_{AC}+\hat
{V}_{BC}+\hat{V}_{ABC},
\]
where $\hat{H}_{A},\hat{H}_{B},\hat{H}_{C}$ are free Hamiltonians for the late
radiation, the black hole and the early radiation, and $\hat{V}_{AB},\hat
{V}_{AC},\hat{V}_{BC}$ are two-body interactions among $A,B,C$, and $\hat
{V}_{ABC}$ is a three-body interaction (if we have). The setup is depicted in
figure 5. Let us think that $\hat{V}_{AC}+\hat{V}_{BC}+\hat{V}_{ABC}$ are
negligibly small as usual in statistical mechanics setups. However, we do not
neccesalily assume that $\hat{V}_{AB}$ is small. Then $\overline{\hat{\rho
}_{AB}}$ does not take the form of eq. (\ref{4}), but instead%
\begin{equation}
\overline{\hat{\rho}_{AB}}=\frac{\exp\left(  -\beta\left(  \hat{H}_{A}+\hat
{H}_{B}+\hat{V}_{AB}\right)  \right)  }{\operatorname*{Tr}_{AB}\left[
\exp\left(  -\beta\left(  \hat{H}_{A}+\hat{H}_{B}+\hat{V}_{AB}\right)
\right)  \right]  }. \label{5}%
\end{equation}
The expression of eq. (\ref{5}) is correct irrespective of the interaction
strength between $A$ and $B$. The expectation value of $\hat{V}_{AB}$, which
includes the kinetic energy term of radiation fields on the horizon, does not
diverge at all:%
\[
\left\vert \operatorname*{Tr}_{AB}\left[  \overline{\hat{\rho}_{AB}}\hat
{V}_{AB}\right]  \right\vert <\infty.
\]
It can be easily understood if we notice that the decomposition of $AB$ system
into two parts ($A$ and $B$) is arbitrary. If we choose the boundary of two
systems differently, we have different subsystems $A^{\prime}$ and $B^{\prime
}$ and different free Hamiltonians $\hat{H}_{A^{\prime}}^{\prime},\hat
{H}_{B^{\prime}}^{\prime}~$and interaction $\hat{V}_{A^{\prime}B^{\prime}%
}^{\prime}$ between them. But physics of the composite system does not change
at all because%

\[
\hat{H}_{A}+\hat{H}_{B}+\hat{V}_{AB}=\hat{H}_{A^{\prime}}^{\prime}+\hat
{H}_{B^{\prime}}^{\prime}+\hat{V}_{A^{\prime}B^{\prime}}^{\prime}%
\]
holds in eq. (\ref{5}). From the viewpoint of $A^{\prime}$ and $B^{\prime}$,
$\hat{V}_{AB}$ is an ordinary local operator of $A^{\prime}$ or $B^{\prime}$.
There is no cause to make $\operatorname*{Tr}_{AB}\left[  \overline{\hat{\rho
}_{AB}}\hat{V}_{AB}\right]  \left(  =\operatorname*{Tr}_{A^{\prime}B^{\prime}%
}\left[  \overline{\hat{\rho}_{A^{\prime}B^{\prime}}}\hat{V}_{AB}\right]
\right)  $ diverge. This remains true even if we take the limit of $\hat
{V}_{AB}\rightarrow0$ and recover the expression in eq. (\ref{4}). After all,
we have no reasoning for firewall emergence on the horizon from a viewpoint of
quantum information. \begin{figure}[ptb]
\begin{center}
\includegraphics[width=12cm]{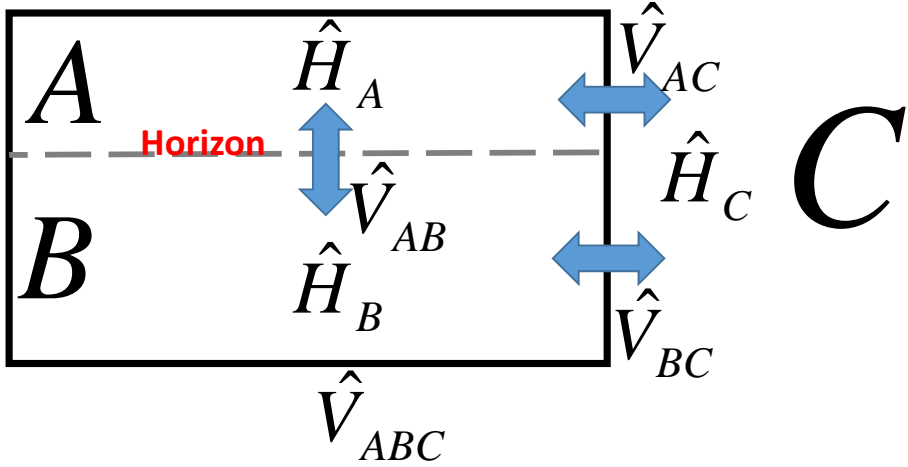}
\end{center}
\caption{a canonical typicality setup for an old black hole, early radiation
and late radiation, which allows all energy exchange among them, preserving
total energy.}%
\end{figure}

Here it is worth emphasing that the LLPP theorem can be regarded as a special
case of the canonical typicality. When we consider a nondegenerate Hamiltonian
with a cut off, the density of states $e^{S(E)}$ has the maximum value at an
energy close to the cut off (See figure 3), and the temperature is infinite at
this point since $T^{-1}=\beta=\frac{\partial S}{\partial E}=0$. If we choose
a state randomly from the whole Hilbert space without any energy condition, we
almost always get a state with energy corresponding to the maximum density of
states. Then the subsystem has the Gibbs state with $\beta=0$, which is
proportional to the identity operator. This is exactly what the LLPP theorem
claims. Consequently, we obtain a very high energy state with infinite
temperature. Therefore the firewall argument based on the LLPP theorem
actually says that not only the horizon, but also the whole space is on fire.
(Note that the argument in \cite{firewall} does not use any special property
of the horizon. Therefore it can be applied to an arbitrary partition of the
space.) This wrong conclusion teaches us the importance of introducing a
physical Hamiltonian and the energy conservation law to consider this problem.

\bigskip

\bigskip

\section{Possible Modification of Page Curve}

\bigskip

In this section, we revisit Page's proposition (II). In section 2, we
explained that quantum state $\hat{\rho}_{1}$ of the smaller system $S_{1}$ in
a typical state $|\Psi\rangle_{12}$ of $\mathcal{H}_{\Delta(E)}$ \ equals a
Gibbs thermal state. Because entanglement entropy is defined as $S_{EE}%
=-\operatorname*{Tr}_{1}\left[  \hat{\rho}_{1}\ln\hat{\rho}_{1}\right]  $,
\ $S_{EE}$ is actually the same as thermal entropy when $S_{1}~$and $S_{2}$
exchanges energy via the boundary, and interaction $\hat{V}_{12}$ is
negligibly small compared to the free Hamiltonians. This condition of small
$\hat{V}_{12}$ really holds in nearest-neighbor interaction cases, because the
free Hamiltonians are proportional to volumes of $S_{1}~$and $S_{2}$, and
$\hat{V}_{12}$ is merely proportional to the boundary area. Thus proposition
(II) might sound convincing. However, we should not forget a crucial condition
which makes $\hat{\rho}_{1}$ a typical state. Energy has to be transported not
only from $S_{1}$ to $S_{2}$, but also from $S_{2}$ to $S_{1}$. In such a
situation, all of energy eigenstates in $\mathcal{H}_{\Delta(E)}$ are able to
contribute on the same footing with each other, as depicted in figure 6.
\begin{figure}[ptb]
\begin{center}
\includegraphics[width=8cm]{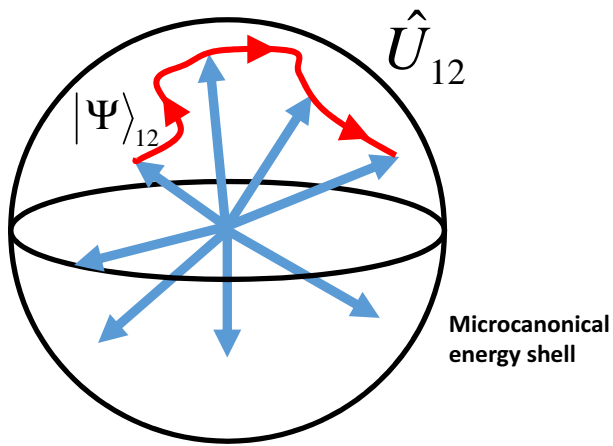}
\end{center}
\caption{fast scrambling into typical states uniformly distributed all over
the energy shell.}%
\end{figure}$\hat{U}_{12}$ is expected to generate very complicated time
evolution, and may yield fast scrambling of the system in $\mathcal{H}%
_{\Delta(E)}$. Here 'scrambling' means relaxation of nontypical initial states
with zero entanglement into typical states with high entanglement. After the
relaxation, it is very unlikely to find the system in a nontypical state again
for ordinary systems. This makes the canonical typicality method in section 2
promising for late time. Finding typical $\hat{\rho}_{1}$ makes sense after
the relaxation. How about cases in which energy is transferred only from
$S_{1}$ to $S_{2}$? In these cases, the energy transportation does not happen
from $S_{2}$ to $S_{1}$, as depicted in figure 7. \begin{figure}[ptb]
\begin{center}
\includegraphics[width=12cm]{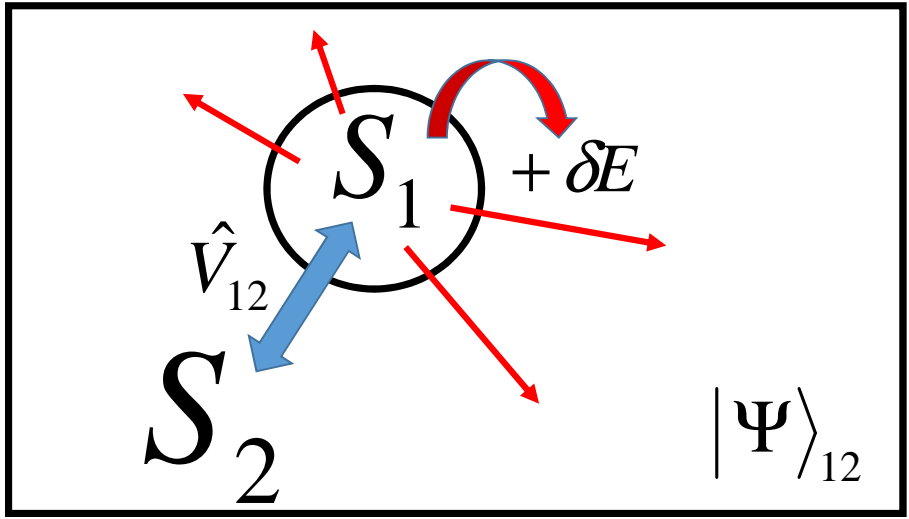}
\end{center}
\caption{an actual setup for black hole evaporation, in which energy is
transferred only from the black hole to radiation. The energy transportation
from outgoing Hawking radiation to the black hole does not take place. }%
\end{figure}This actually arises in black hole evaporation, because the
outgoing Hawking radiation emitted by black holes does not come back. The
radiation is not able to give any amount of energy to the black holes without
putting a mirror outside the horizon, or assuming thermal equilibrium. As well
as the negative heat capacity of evaporating black holes, the one way energy
transportation is caused by the gravitational instability. Such a one-way
dynamics of energy transportation makes the system remain in nontypical states
before the last burst of the black hole. Generation of the Hawking radiation
takes place in an outside region (a few times the black hole radius far from
the horizon) with very small spacetime curvature. In an ordinary sense, the
semi-classical treatment of the generation is justified, and the process is
not random at all. Time evolution of $B$ and $R$ is described essentially by
fast scrambling $\hat{U}_{B}\otimes\hat{I}_{R}$ of black holes and nonrandom
process $\hat{U}_{BR}^{(emission)}$ of the Hawking radiation emission.
$\hat{U}_{B}\otimes\hat{I}_{R}$ does not change entanglement between $B$ and
$R$ due to its locality. No fast scrambling for $R$ happens. This nonchaotic
setup does not ensure that the relaxation of $BR$ system finishes before the
last burst. Therefore, in realistic evaporations, the validity of the typical
state postulate becomes dubious. The system may always evolve among nontypical
states as depicted in figure 8. \begin{figure}[ptb]
\begin{center}
\includegraphics[width=8cm]{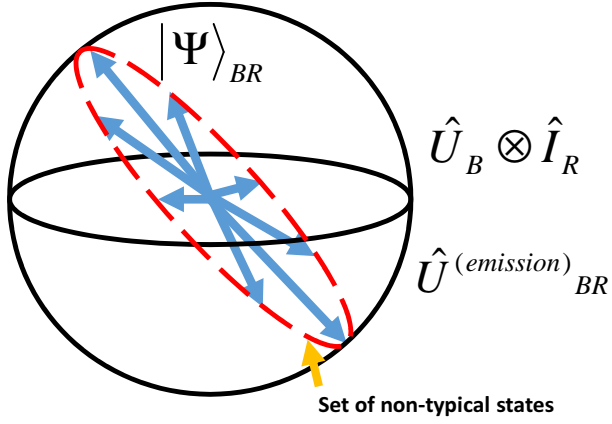}
\end{center}
\caption{a slow relaxation into typical states, which is generated by
fast-scrambling inside the black hole and semi-classical radiation emission. }%
\end{figure}Therefore the quantum state of the old black hole is able to be
far from Gibbs states after the Page time. Hence $S_{EE}$ can be totally
different from thermal entropy, as opposed to Page's proposition (II).

Taking account of the possibility without (II), it is of significant to
discuss modification of the Page curve. Moving mirror models, which mimic
gravitational collapse and Hawking radiation emission out of black holes, may
be a good device to see the possibilities. Let us consider a massless scalar
field in 1+1 dimensions. Adopting light cone coordinates $x^{\pm}=t\pm x$, the
mirror trajectory is expressed as
\[
x^{+}=f\left(  x^{-}\right)  ,
\]
where $f$ is a monotonically increasing function of $x^{-}$. Even if the
vacuum state is set as the initial state of the field, the mirror excites the
field and emits radiation, whose expectation value of the outgoing energy flux
\bigskip is computed \cite{BD} as%
\begin{equation}
\left\langle \hat{T}_{--}(x^{-})\right\rangle =-\frac{1}{24\pi}\left[
\frac{\partial_{x^{-}}^{3}f(x^{-})}{\partial_{x^{-}}f(x^{-})}-\frac{3}%
{2}\left(  \frac{\partial_{x^{-}}^{2}f(x^{-})}{\partial_{x^{-}}f(x^{-}%
)}\right)  ^{2}\right]  . \label{13}%
\end{equation}
A trajectory given by%
\begin{equation}
f_{o}(x^{-})=-\ln\left(  1+e^{-\kappa x^{-}}\right)  \label{14}%
\end{equation}
describes a mirror which is at rest in the past, and accelerates with constant
acceleration $\kappa$ in the future. This trajectory is related to a realistic
1+3 dimensional gravitational collapse, which makes an eternal black hole
without the back reaction of radiation emission \cite{w}. Eventually the
mirror emits constant thermal flux with temperature $T=\kappa/(2\pi)$. In
fact, substitution of eq. (\ref{14}) into eq. (\ref{13}) yields the correct
thermal flux,
\[
\left\langle \hat{T}_{--}(x^{-})\right\rangle =\frac{\pi}{12}T^{2}%
\]
for $x^{-}\gg1/\kappa$. Now let us think mirror trajectories which may
approximately describe black hole evaporation with its back reaction. The
first candidate is the following:
\begin{equation}
f_{\kappa}(x^{-})=-\ln\left(  \frac{1+e^{-\kappa x^{-}}}{1+e^{\kappa\left(
x^{-}-h\right)  }}\right)  , \label{15}%
\end{equation}
where $h$ is a very large real constant, and controls the lifetime of the
corresponding black hole. The trajectory is depicted in figure 9. Due to the
trajectory deformation, the mirror stops in the future. The time evolution of
radiation emission is given by the plot in figure 10 for $\kappa=1$ and
$h=500$. During the evaporation, almost constant flux is emitted, though real
black holes increase the temperature and flux of radiation. It is interesting
to compute entanglement entropy $S_{EE}$ between the field degrees of freedom
inside $\left[  x_{1}^{-},x_{2}^{-}\right]  $ and those outside $\left[
x_{1}^{-},x_{2}^{-}\right]  $. There exists a ultraviolet divergence in
$S_{EE}$ due to infinite number of degrees of freedom of the quantum field
\cite{sr}. To remove the divergence, a renormalized entanglement entropy
$\Delta S_{EE}$ is introduced by substituting the vacuum contribution
\cite{hlw}. $\Delta S_{EE}$ is given by%

\[
\Delta S_{EE}=\frac{1}{12}\ln\left(  \frac{\left(  f(x_{2}^{-})-f(x_{1}%
^{-})\right)  ^{2}}{\left(  x_{2}^{-}-x_{1}^{-}\right)  ^{2}\partial_{x^{-}%
}f(x_{2}^{-})\partial_{x^{-}}f(x_{1}^{-})}\right)  .
\]
In figure 11, a plot of $\Delta S_{EE}$ as a function of $x_{2}^{-}$ is
provided for $\kappa=1$, $h=500$, and $x_{1}^{-}=-2$. The curve is almost
symmetric and looks like the Page curve \cite{t to H}. This aspect comes from
a fact that the order of the mirror deceleration, which is expected due to
back reaction of the radiation emission, is equal to $\kappa$ of the
acceleration in eq. (\ref{15}). However, this means locality breaking of
dynamics for the evaporation and seems unlikely. $\kappa$ corresponds to the
scale of surface gravity of the 1+3 dimensional black hole, and is considered
as the order of $M_{pl}^{2}/M_{BH}$, where $M_{pl}$ is Planck mass and
$M_{BH}$ is black hole mass. Thus $\kappa$ is very small compared to $M_{pl}$
and the inverse of $\kappa$ provides a cosmologically long time scale. The
macroscopic black hole, whose evolution is described by eq. (\ref{15}), needs
to estimate by itself its destiny, how much time remains before its death, and
when the deceleration must start. At the half of lifetime ($x^{-}\sim250$),
the black hole decides to emit its quantum information, which is stored inside
the horizon, so as to finish leaking all the information before the last
burst. In order to achieve this, the black hole has to slightly change its
evolution at its Page time, which is much before its death, in a different way
from those of other black holes with the same mass. For example, let us think
two black holes with the same mass $M_{BH}$, as depicted in figure 12. The
left black hole in the figure is just born and very young. It does not begin
the emission of the Hawking radiation yet and is almost in a pure state. The
right black hole in figure 12 was born with its mass $1.3M_{BH}~(\sim
M_{BH}/0.77)$, and has decreased the mass to $M_{BH}$ via the radiation
emission. Thus it is an old black hole around the Page time. It is worth
stressing that classical geometries of the two macroscopic black holes are the
same. Nevertheless, only the old black hole begins to change its evolution at
$x^{-}\sim250$. The young black hole will change in a similar way in a further
future. To perform such a cooperative motion, all quantum microscopic
ingredients of the black hole must watch each other carefully and preserve the
long-term memory. Each part must estimate the black hole age by use of the
memory. This dynamics is non-Markovian and non-local at least in time. If
their dynamics can be approximated by the semi-classical general relativity,
which is Markovian, such nonlocal evolution does not emerge. One might expect
that small back reaction of the Hawking radiation to the black hole geometry
becomes a trigger to leak the information for the old black hole. However, if
the black hole has such a sensitivity, the geometrical perturbation induced by
a small amount of infalling matter must also drastically change the time
schedule of information leakage and completely modify the Page curve. This
implies that Page curve is unstable and useless for realistic situations.
\begin{figure}[ptb]
\begin{center}
\includegraphics[width=8cm]{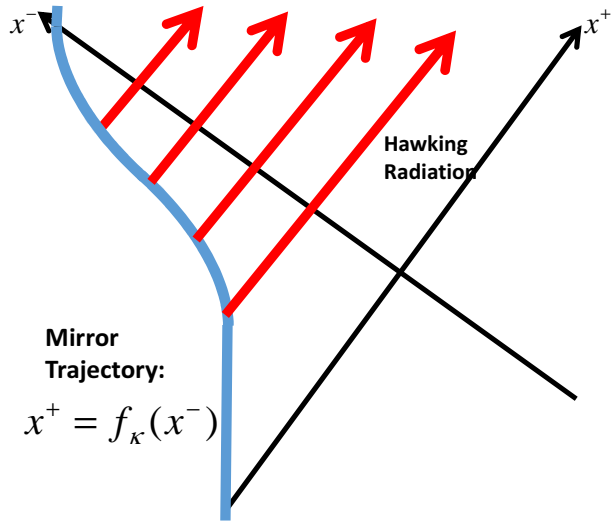}
\end{center}
\caption{schematic figure of mirror trajectory described by eq. (\ref{15}).}%
\end{figure}\begin{figure}[ptb]
\begin{center}
\includegraphics[width=12cm]{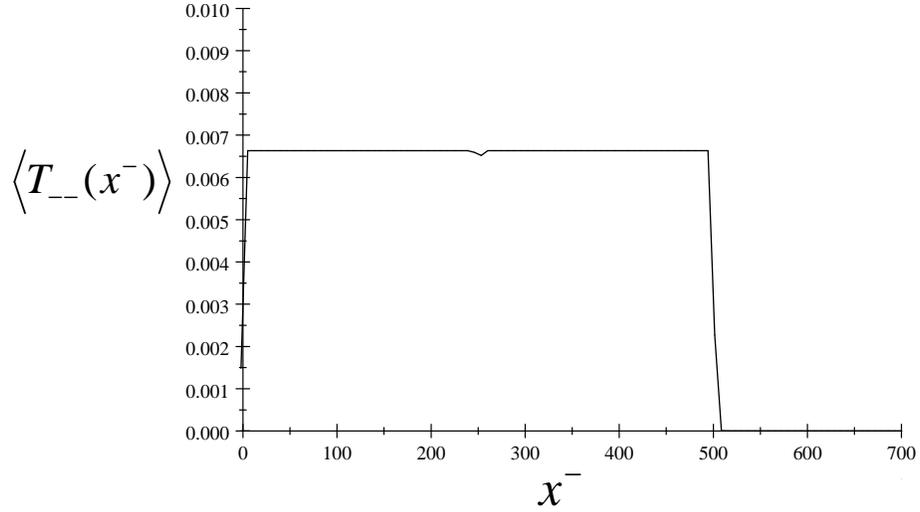}
\end{center}
\caption{energy flux of Hawking radiation out of a mirror for the trajectory
of eq. (\ref{15}) with $\kappa=1$ and $h=500$. }%
\end{figure}\begin{figure}[ptb]
\begin{center}
\includegraphics[width=12cm]{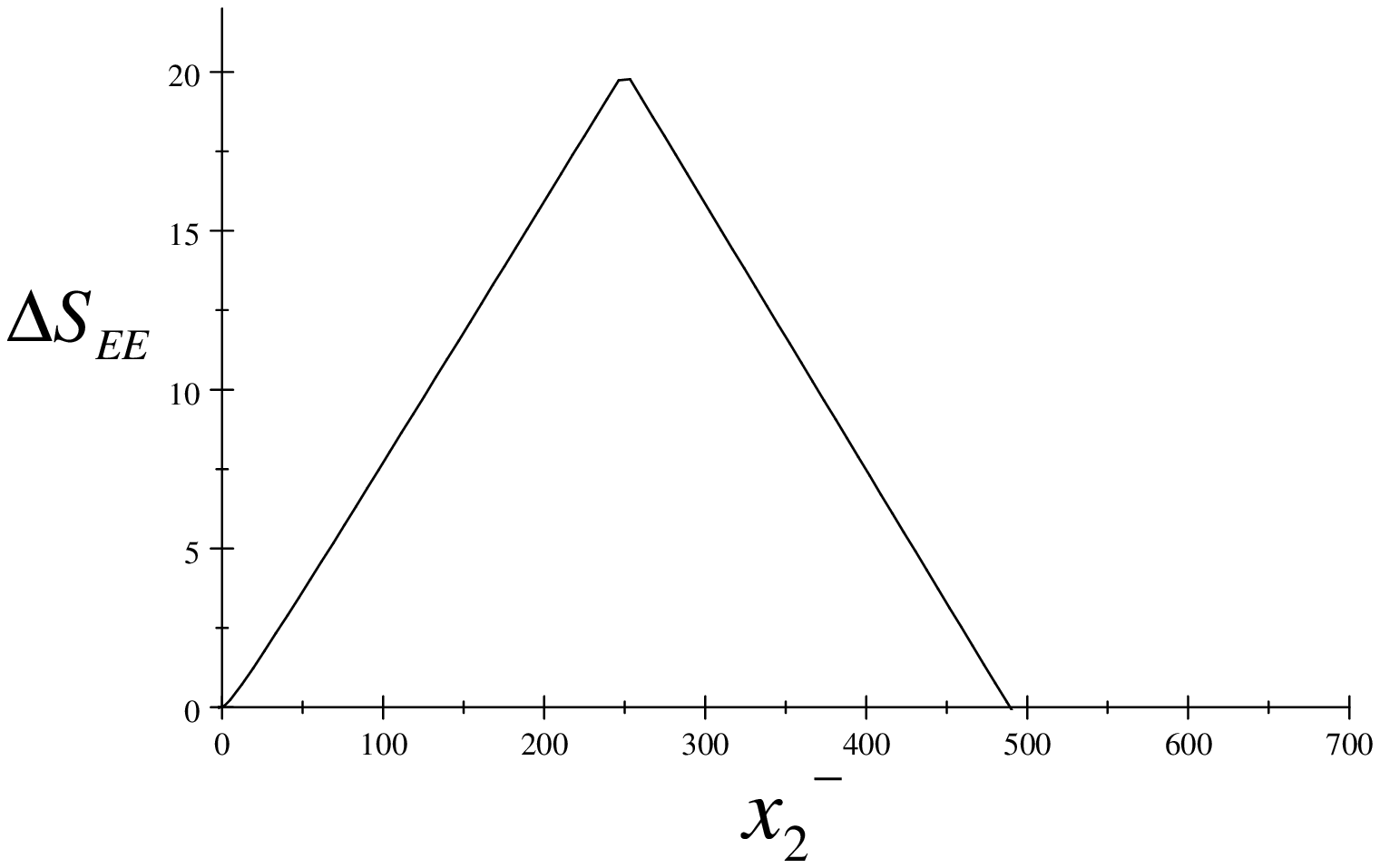}
\end{center}
\caption{time curve of renormalized entanglement entropy between degrees of
freedom in $\left[  -2,x_{2}\right]  $ and outside ones for the trajectory of
eq. (\ref{15}) with $\kappa=1$ and $h=500$. }%
\end{figure}\begin{figure}[ptb]
\begin{center}
\includegraphics[width=12cm]{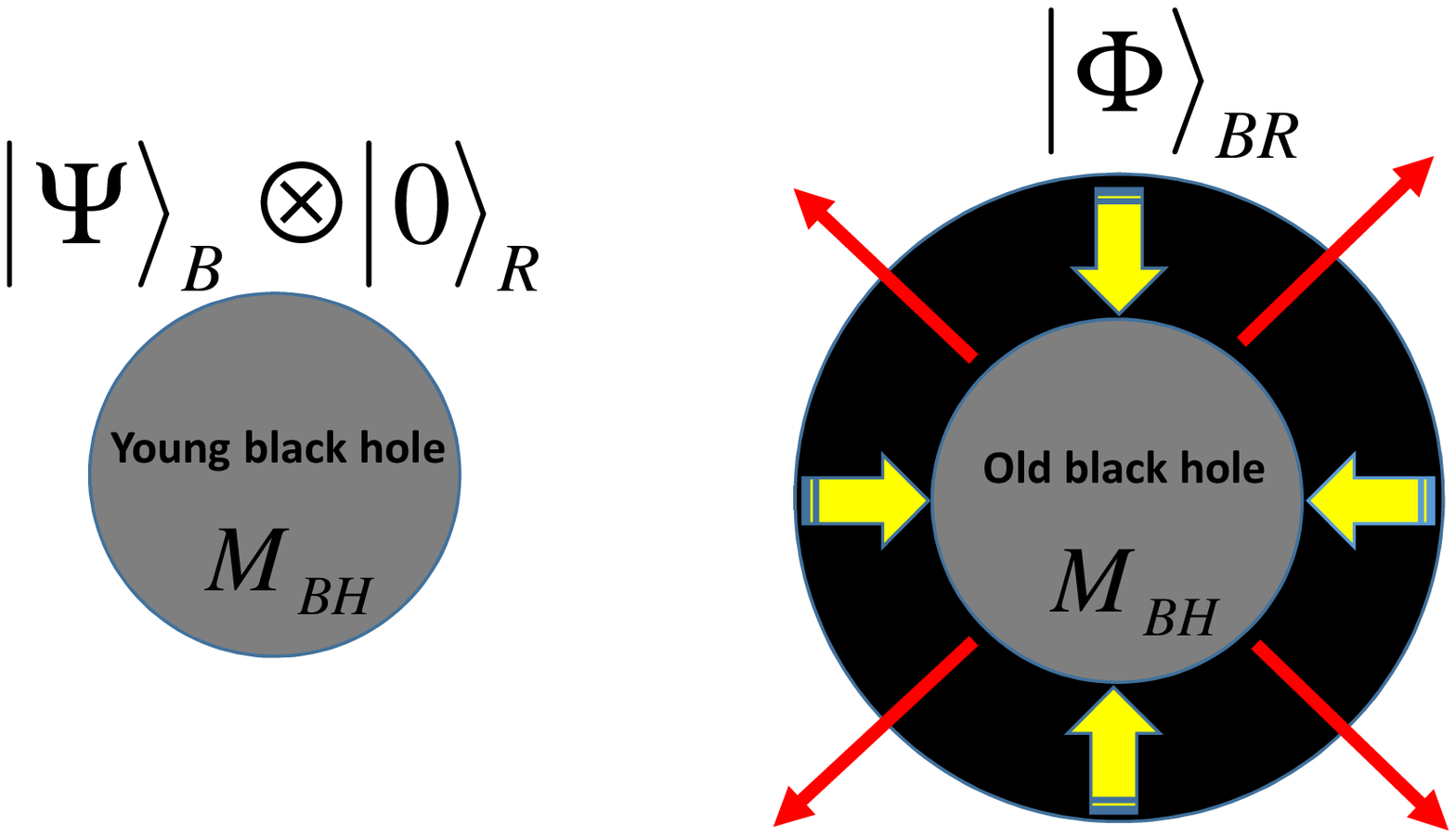}
\end{center}
\caption{young and old black holes with the same mass. The entanglement
evolutions are quite different from each other in Page curve hypothesis. }%
\end{figure}

Of course, we are able to think a more natural mirror trajectory, which is
consistent with the semi-classical general relativity. One of the examples is
given by
\begin{equation}
f_{\kappa,\lambda}(x^{-})=-\ln\left(  \frac{1+e^{-\kappa x^{-}}}%
{1+e^{\lambda\left(  x^{-}-h\right)  }}\right)  . \label{16}%
\end{equation}
Here two different scale parameters $\kappa,\lambda$ are introduced. $\kappa$
is the acceleration parameter for the emission of Hawking radiation. $\lambda$
is the deceleration parameter, which is of Planck scale order $M_{pl}$ and
describes the sudden stop of the mirror due to the last burst of the black
hole. Thus $\lambda\gg\kappa$ holds. The schematic behavior is given in figure
13. Clearly the last burst region should be described by quantum gravity. In
figure 14, $\Delta S_{EE}$ is plotted as a function of $x_{2}^{-}$ for
$\kappa=1$, $\lambda=100,$ $h=500$, and $x_{1}^{-}=-2$. Note that, until just
before the last burst, $\Delta S_{EE}$ is estimated by the semi-classical
results of the outside Hawking radiation without use of knowledge about
quantum black holes. This aspect circumvents uncertainty of quantum gravity,
and strengthens the plausibility of this scenario. All the information is
retrieved by the last burst. It is often commented that a huge amount of
energy is required for such a information leakage However, it may be possible
to attain it by use of entanglement with zero-point fluctuating of quantum
fields \cite{w}\cite{HMF}\cite{HSU} as already emphasized above. Hence the
energy of Planck mass order is enough to leak the information. In order to see
the possibility, let us think an entangled particle pair in the vacuum state
\cite{HSU}, as depicted in figure 15. One of the particles becomes a Hawking
particle with positive energy after scattering by the mirror. However the
partner particle is scattered by the mirror at rest and has no energy even
after the scattering. Thus an enormous amount of quantum information, which
makes the system state pure, is shared among the Hawking radiation and the
outgoing zero-point fluctuation flow of quantum fields in the future null
infinity. When this scenario is applied to realistic black hole evaporation,
we would expect that quantum gravity generates higher derivative corrections
to Einstein equation and the smeared would-be singularity inside the black
hole horizon becomes timelike and preserves the information of falling matters.

Here we add some comments. The famous qualitative bound for the information
leakage time of Carlitz and Willey \cite{remnant2} is unable to be applied to
the sudden stop case in figure 15, because their bound is derived by assuming
slow change of acceleration of the mirror and ignoring emission of negative
energy flux, which is generated with the informational zero-point fluctuaion.
Recently Bianchi and Smerlak found an interesting identity between the energy
flux emitted from moving mirrors and the entanglement entropy of the radiation
\cite{BS}. They also argue that when applied to two-dimensional models of
black hole evaporation, this identity implies that unitarity is incompatible
with monotonic mass loss. However, it should be stressed that they assume the
Page curve as a unitary model of black hole evaporation for the argument. Thus
the outgoing zero-point fluctuation flow scenario, as well as the long-lived
remnant scenario, is consistence with their claim. It seems natural that the
singluar sudden stop of mirror trajectory in figure 15 may be smoothed around
Planck length scale by quantum gravity, as depicted in figure 16. Then a
gravitational shock wave induced by the last burst ray may trap the entangled
partner particle with zero energy for a while, just like in a high-energy
gravitational scattering in a Minkowski vacuum. Even after reemission of the
particle out of the shock wave, the energy of the partner particle may remain
zero \cite{gsw}. Thus the possibility is still alive that the information
leakage by the last burst does not need a huge amount of energy in quantum
gravity. This totally differs from the Page curve hypothesis, but is one of
interesting possibilities. \begin{figure}[ptb]
\begin{center}
\includegraphics[width=8cm]{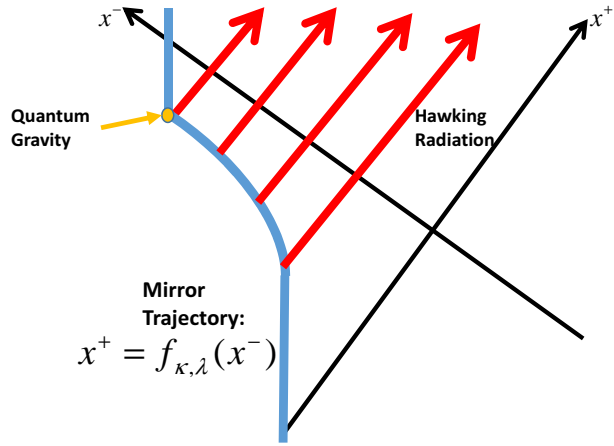}
\end{center}
\caption{schematic figure of mirror trajectory described by eq. (\ref{16}).}%
\end{figure}\begin{figure}[ptb]
\begin{center}
\includegraphics[width=11cm]{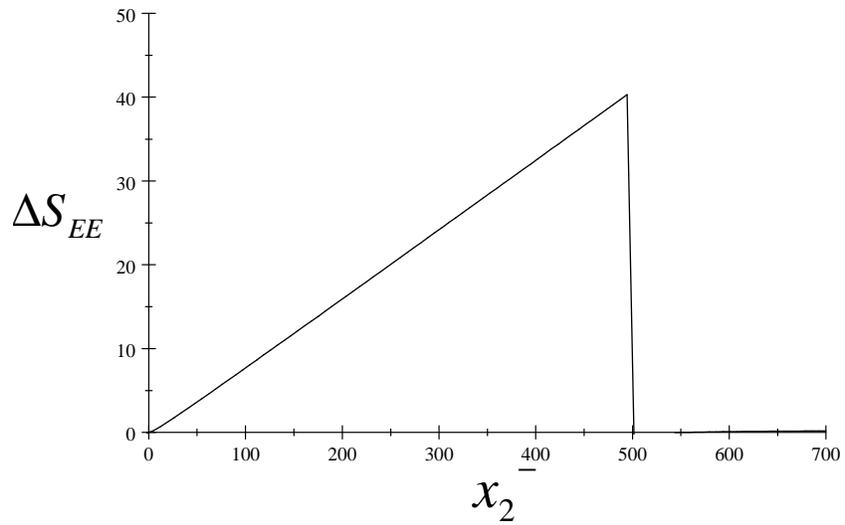}
\end{center}
\caption{time curve of renormalized entanglement entropy between degrees of
freedom in $\left[  -2,x_{2}\right]  $ and outside ones for the trajectory of
eq. (\ref{16}) with $\kappa=1,\lambda=100$ and $h=500$. }%
\end{figure}

Before closing this section, we add an interesting comment about a thermal
equilibrium for $BR$ system. Instead of black hole evaporation, let us suppose
a pure state which describes a static and stable thermal equilibrium of the
composite system in a coarse-grained meaning \cite{pts}. In asymptotically
anti-de Sitter spacetimes, there exist thermal equilbriums for a large black
hole $B$ and its Hawking radiation $R$. They exchange energy bi-directionally.
By adiabatically slowly changing system parameters including external forces,
and position and pressure of a mirror surrounding $R$ (if we have the mirror),
various sizes of the black hole may appear. From the general results in
section 2, it turns out that each equilibrium state is typical, and the
reduced state for the smaller subsystem among $B$ and $R$ is a Gibbs state
with finite temperature, though no firewall emerges. Plotting entanglement
entropy as a function of the inverse of black hole size generates a Page-like
curve, in which entanglement entropy equals thermal entropy for the smaller
subsystem. This may become relevant in the future research of quantum black
holes, though it is merely a side story for the original information loss
problem. \begin{figure}[ptb]
\begin{center}
\includegraphics[width=12cm]{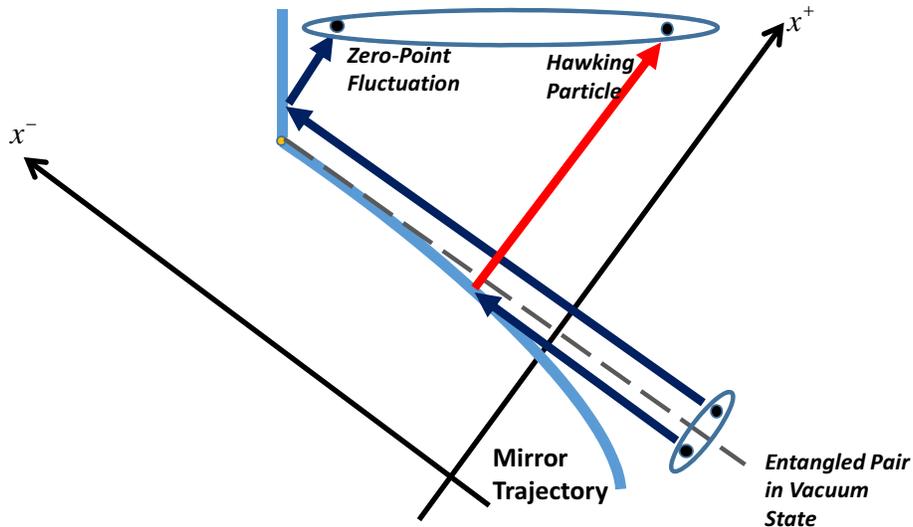}
\end{center}
\caption{entangled particles scattered by mirror.}%
\end{figure}\begin{figure}[ptb]
\begin{center}
\includegraphics[width=12cm]{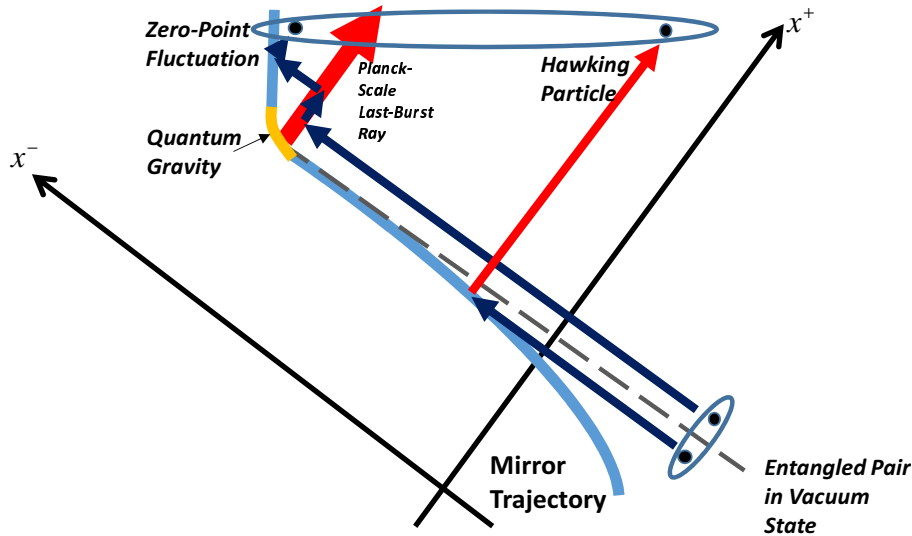}
\end{center}
\caption{gravitational shock wave induced by the last burst ray and spacetime
shift of zero-point fluctuation.}%
\end{figure}

\newpage

\bigskip

\bigskip

\section{Summary}

\bigskip

\ ~~

In this paper, we revisit the Page curve hypothesis. Adopting a general
formulation of canonical typicality with nondegenerate Hamiltonian, it is
proven that Page's proposition (I) is not actually satisfied for ordinary
systems. The typical states are exponentially close to Gibbs states with
finite temperatures. The entanglement between subsystems becomes nonmaximal
and removes firewalls. The microcanonical state, which is proportional to
$\hat{I}_{E}$, is far from typical for the entanglement between a black hole
and its Hawking radiation. In the dynamical situation of black hole
evaporation, proposition (II) is also unlikely. We have no strong reason to
expect that the entanglement entropy equals the thermal entropy for the
smaller subsystem in the evaporation. Taking account of the semi-classical
general relativity, a conservative scenario becomes more fascinating in which
all the information inside the black hole is emitted by the last burst of the
black hole. Finally, it is pointed out, using the general results in section
2, that for static thermal pure states of $BR$ system in the sense of
canonical typicality, entanglement entropy between $B$ and $R$ certainly
coincides with thermal entropy of the smaller system. This holds for large AdS
black holes.

\bigskip

\bigskip

\textbf{Acknowledgments}\newline

We acknowledge support from the workshops, "Relativistic Quantum Information
North 2014" at University of Seoul and "Molecule-Type Workshop on Black Hole
Information Loss Paradox" at YITP in Kyoto. We would like to thank Bill Unruh,
Don Page, Daniel Harlow, Yasusada Nambu, Misao Sasaki, Pijin Chen, Anzhong
Wang, Yen Chin Ong, Dong-hang Yeom, Shinji Mukohyama, Ted Jacobson for related
discussions. In particular, we appreciate long and sincere discussion with
Page and private communication with Harlow.

\bigskip

Let us take a uniform distribution in the microcanonical energy shell
$\mathcal{H}_{\Delta(E)}$ as%

\[
p\left(  c_{1},\cdots,c_{D}\right)  =\frac{\Gamma\left(  D-\frac{1}{2}\right)
}{2\pi^{D-1/2}}\delta\left(  \sum_{j\in\Delta(E)}\left\vert c_{j}\right\vert
^{2}-1\right)
\]
and introduce ensemble average of a function of $c_{j}$ as%
\[
\overline{f\left(  c_{1},\cdots,c_{D}\right)  }=\int f\left(  c_{1}%
,\cdots,c_{D}\right)  p\left(  c_{1},\cdots,c_{D}\right)  d^{D}c.
\]
Using invariant tensors of $U(D)$, it is easy to show%
\begin{align}
\overline{c_{j}^{\ast}c_{j^{\prime}}}  &  =\frac{1}{D}\delta_{jj^{\prime}%
},\label{9}\\
\overline{c_{j}^{\ast}c_{j^{\prime}}c_{k}^{\ast}c_{k^{\prime}}}  &  =\frac
{1}{D(D+1)}\left(  \delta_{jj^{\prime}}\delta_{kk^{\prime}}+\delta
_{jk^{\prime}}\delta_{kj^{\prime}}\right)  . \label{10}%
\end{align}
A pure state of $\mathcal{H}_{\Delta(E)}$ is given by
\[
\hat{\rho}\left(  c_{1},\cdots,c_{D}\right)  =|\Psi\rangle\langle\Psi
|=\sum_{j,j^{\prime}\in\Delta(E)}c_{j}c_{j^{\prime}}^{\ast}|E_{j}%
\rangle\langle E_{j^{\prime}}|.
\]
Using eq. (\ref{9}), the ensemble average of expectation value of observable
$\hat{O}$ is computed as%
\[
\overline{\left\langle \hat{O}\left(  c_{1},\cdots,c_{D}\right)  \right\rangle
}=\frac{1}{D}\sum_{j\in\Delta(E)}\langle E_{j}|\hat{O}|E_{j}\rangle.
\]
For the ensemble deviation,
\[
\delta\left\langle \hat{O}\left(  c_{1},\cdots,c_{D}\right)  \right\rangle
=\left\langle \hat{O}\left(  c_{1},\cdots,c_{D}\right)  \right\rangle
-\overline{\left\langle \hat{O}\left(  c_{1},\cdots,c_{D}\right)
\right\rangle },
\]
the mean square error is given by%
\[
\overline{\left(  \delta\left\langle \hat{O}\left(  c_{1},\cdots,c_{D}\right)
\right\rangle \right)  ^{2}}=\overline{\left\langle \hat{O}\left(
c_{1},\cdots,c_{D}\right)  \right\rangle ^{2}}-\left(  \overline{\left\langle
\hat{O}\left(  c_{1},\cdots,c_{D}\right)  \right\rangle }\right)  ^{2}.
\]
Using eq. (\ref{10}), we have%
\begin{align*}
\overline{\left\langle \hat{O}\left(  c_{1},\cdots,c_{D}\right)  \right\rangle
^{2}}  &  =\sum_{j,j^{\prime}\in\Delta(E)}\sum_{k,k^{\prime}\in\Delta
(E)}\overline{c_{j}^{\ast}c_{j^{\prime}}c_{k}^{\ast}c_{k^{\prime}}}\langle
E_{j}|\hat{O}|E_{j^{\prime}}\rangle\langle E_{k}|\hat{O}|E_{k^{\prime}}%
\rangle\\
&  =\frac{1}{D(D+1)}\sum_{j,k\in\Delta(E)}\left(  \langle E_{j}|\hat{O}%
|E_{j}\rangle\langle E_{k}|\hat{O}|E_{k}\rangle+\langle E_{j}|\hat{O}%
|E_{k}\rangle\langle E_{k}|\hat{O}|E_{j}\rangle\right) \\
&  =\frac{D}{D+1}\overline{\left\langle \hat{O}\left(  c_{1},\cdots
,c_{D}\right)  \right\rangle }^{2}+\frac{1}{D(D+1)}\sum_{j,k\in\Delta
(E)}\left\vert \langle E_{j}|\hat{O}|E_{k}\rangle\right\vert ^{2}.
\end{align*}
Thus the mean square error is estimated as%
\[
\overline{\left(  \delta\left\langle \hat{O}\left(  c_{1},\cdots,c_{D}\right)
\right\rangle \right)  ^{2}}=\frac{1}{D(D+1)}\sum_{j,k\in\Delta(E)}\left\vert
\langle E_{j}|\hat{O}|E_{k}\rangle\right\vert ^{2}-\frac{1}{D^{2}\left(
D+1\right)  }\left(  \sum_{j\in\Delta(E)}\langle E_{j}|\hat{O}|E_{j}%
\rangle\right)  ^{2}%
\]
Because $\left(  \sum_{j\in\Delta(E)}\langle E_{j}|\hat{O}|E_{j}%
\rangle\right)  ^{2}$ and $\sum_{k\notin\Delta(E)}\left\vert \langle
E_{j}|\hat{O}|E_{k}\rangle\right\vert ^{2}$are nonnegative, we get%
\begin{align}
\overline{\left(  \delta\left\langle \hat{O}\left(  c_{1},\cdots,c_{D}\right)
\right\rangle \right)  ^{2}}  &  \leq\frac{1}{D(D+1)}\sum_{j\in\Delta(E)}%
\sum_{k\in\Delta(E)}\left\vert \langle E_{j}|\hat{O}|E_{k}\rangle\right\vert
^{2}\nonumber\\
&  \leq\frac{1}{D(D+1)}\sum_{j\in\Delta(E)}\sum_{k=1}^{D_{1} D_{2}}\left\vert
\langle E_{j}|\hat{O}|E_{k}\rangle\right\vert ^{2}\\
&  =\frac{1}{D(D+1)}\sum_{j\in\Delta(E)}\langle E_{j}|\hat{O}^{2}|E_{j}%
\rangle\nonumber\\
&  =\frac{1}{D+1}\overline{\left\langle \hat{O}\left(  c_{1},\cdots
,c_{D}\right)  ^{2}\right\rangle } \label{12}%
\end{align}
Here we have used%
\[
\sum_{k=1}^{D_{1}D_{2}}\left\vert \langle E_{j}|\hat{O}|E_{k}\rangle
\right\vert ^{2}=\langle E_{j}|\hat{O}\left(  \sum_{k=1}^{D_{1}D_{2}}%
|E_{k}\rangle\langle E_{k}|\right)  \hat{O}|E_{j}\rangle=\langle E_{j}|\hat
{O}^{2}|E_{j}\rangle.
\]

Because an expectation value does not exceed its operator norm,
\begin{equation}
\overline{\left\langle \hat{O}\left(  c_{1},\cdots,c_{D}\right)
^{2}\right\rangle }\leq\left\Vert \hat{O}^{2}\right\Vert \label{11}%
\end{equation}
is always satisfied. Combining eq. (\ref{12}) and eq. (\ref{11}) yields eq.
(\ref{8}).

The ensemble average state for $S_{1}$ is given by%

\[
\overline{\hat{\rho}_{1}}=\frac{1}{D_{1}}\left(  \hat{I}+\sum_{n}%
\overline{\left\langle \hat{G}_{n0}\left(  c_{1},\cdots,c_{D}\right)
\right\rangle }\hat{T}_{n}\right)  .
\]
The mean square error is computed as%
\begin{align*}
&  \operatorname*{Tr}_{1}\left[  \left(  \hat{\rho}_{1}\left(  c_{1}%
,\cdots,c_{D}\right)  -\overline{\hat{\rho}_{1}}\right)  ^{2}\right] \\
&  =\frac{1}{D_{1}^{2}}\operatorname*{Tr}_{1}\left[  \left(  \sum_{n}\left(
\left\langle \hat{G}_{n0}\left(  c_{1},\cdots,c_{D}\right)  \right\rangle
-\overline{\left\langle \hat{G}_{n0}\left(  c_{1},\cdots,c_{D}\right)
\right\rangle }\right)  \hat{T}_{n}\right)  ^{2}\right] \\
&  =\frac{1}{D_{1}}\sum_{n}\left(  \left\langle \hat{G}_{n0}\left(
c_{1},\cdots,c_{D}\right)  \right\rangle -\overline{\left\langle \hat{G}%
_{n0}\left(  c_{1},\cdots,c_{D}\right)  \right\rangle }\right)  ^{2}%
\end{align*}
Therefore we can manipulate it as follows.%
\begin{align*}
&  \overline{\operatorname*{Tr}_{1}\left[  \left(  \hat{\rho}_{1}\left(
c_{1},\cdots,c_{D}\right)  -\overline{\hat{\rho}_{1}\left(  c_{1},\cdots
,c_{D}\right)  }\right)  ^{2}\right]  }\\
&  =\frac{1}{D_{1}}\sum_{n}\left(  \overline{\left\langle \hat{G}_{n0}\left(
c_{1},\cdots,c_{D}\right)  \right\rangle ^{2}}-\overline{\left\langle \hat
{G}_{n0}\left(  c_{1},\cdots,c_{D}\right)  \right\rangle }^{2}\right) \\
&  \leq\frac{1}{D_{1}}\sum_{n}\overline{\left(  \delta\left\langle \hat
{G}_{n0}\left(  c_{1},\cdots,c_{D}\right)  \right\rangle \right)  ^{2}}\\
&  =\frac{1}{D_{1}\left(  D+1\right)  }\sum_{n}\left\Vert \hat{G}_{n0}%
^{2}\right\Vert
\end{align*}
In the last step, we used eq. (\ref{8}). Thus eq. (\ref{7}) is proven.

\end{document}